\overfullrule=0pt
\input harvmac
\def\a{{\alpha}}

\def\ad{{\dot a}}
\def\bd{{\dot b}}

\def\l{{\lambda}}
\def\b{{\beta}}

\def\g{{\gamma}}
\def\k{{\kappa}}

\def\d{{\delta}}

\def\e{{\epsilon}}

\def\half{{1\over 2}}
\def\p{{\partial}}
\def\pb{{\bar\partial}}
\def\t{{\theta}}

\Title{\vbox{\hbox{IFT-P.009/2004 }}}
{\vbox{
\centerline{\bf Self-Dual Super-Yang-Mills as a String Theory in
$(x,\theta)$ Space}}}
\bigskip\centerline{Nathan Berkovits\foot{e-mail: nberkovi@ift.unesp.br}
}
\bigskip
\centerline{\it Instituto de F\'\i sica Te\'orica, Universidade Estadual
Paulista}
\centerline{\it Rua Pamplona 145, 01405-900, S\~ao Paulo, SP, Brasil}

\vskip .3in
\centerline{\bf Proceedings from Strings 2003 talk (Kyoto, July 2003)}
\vskip .3in
Different string theories in twistor space have recently been proposed
for describing ${\cal N}=4$ super-Yang-Mills. In this paper, my
Strings 2003 talk is reviewed in which a string theory in $(x,\theta)$
space was constructed for self-dual ${\cal N}=4$ super-Yang-Mills.
It is hoped that these results will be useful for understanding the
twistor-string proposals and their possible relation with the pure
spinor formalism of the $d=10$ superstring.

\Date{March 2004}

\newsec{Motivation}

At the present time, the only quantizable description of the superstring
in an $AdS_5\times S^5$ background uses the pure spinor formalism of
the superstring \ref\pure{N. Berkovits,
{\it 
Super-Poincar\'e Covariant Quantization of the
Superstring}, JHEP 04 (2000) 018, hep-th/0001035\semi
N. Berkovits, {\it ICTP Lectures on Covariant Quantization of the
Superstring}, hep-th/0209059.}\ref\ads{N. Berkovits and
O. Chand\'{\i}a, {\it Superstring Vertex Operators in an
$AdS_5\times S^5$ Background}, Nucl. Phys. B596 (2001) 185,
hep-th/0009168\semi
B.C. Vallilo, {\it One Loop Conformal Invariance of the 
Superstring in an $AdS_5\times S^5$ Background},
JHEP 12 (2002) 042, hep-th/0210064.}.\foot
{Although one can semi-classically describe the
superstring in an $AdS_5\times S^5$ background using the 
covariant Green-Schwarz
formalism \ref\metsaev{R.R. Metsaev and A.A. Tseytlin,
{\it Type IIB Superstring Action in $AdS_5\times S^5$ Background},
Nucl. Phys. B533 (1998) 109, hep-th/9805028.}, 
this description is not fully quantum. Note that even in a flat
background, the covariant Green-Schwarz formalism has not yet been used
to compute superstring scattering amplitudes. Although the light-cone
Green-Schwarz formalism can be quantized in a flat background,
there are
subtleties with quantizing the light-cone gauge formalism in 
an $AdS$ background \ref\thorn{
R.R. Metsaev, C.B. Thorn and A.A. Tseytlin,
{\it Lightcone Superstring in AdS Spacetime}, Nucl. Phys. B596 (2001)
151, hep-th/0009171.}.} When the $AdS$ radius goes to zero, this string
theory is conjectured to describe perturbative ${\cal N}=4$ super-Yang-Mills
\ref\malda{J. Maldacena, {\it The Large N Limit of Superconformal
Field Theories and Supergravity}, Adv. Theor. Math. Phys. 2 (1998) 231,
hep-th/9711200.}.
But because the string theory becomes a strongly coupled sigma model
when the $AdS$ radius goes to zero, it has not yet been possible
to test this conjecture directly using the pure spinor formalism. 
Perhaps the integrable currents found in \ref\BV{B.C. Vallilo,
{\it Flat Currents in the Classical $AdS_5\times S^5$ Pure Spinor 
Superstring}, hep-th/0307018\semi I. Bena, J. Polchinski and R. Roiban,
{\it Hidden Symmetries of the $AdS_5\times S^5$ Superstring}, Phys. Rev.
D69 (2004) 046002, hep-th/0305116.} will be useful for connecting the
large and small radius limits of the string theory. 

Although one cannot yet directly study the small radius limit of the
closed superstring in an $AdS_5\times S^5$ background, it might be possible
to guess the structure of the string theory using the knowledge that
it should describe ${\cal N}=4$ $d=4$ super-Yang-Mills theory.
By analogy with the Gopakumar-Vafa duality \ref\gopa{R. Gopakumar
and C. Vafa, {\it On the Gauge Theory/Geometry Correspondence},
Adv. Theor. Math. Phys. 3 (1999) 1415.} where closed strings on
the conifold are dual to open strings on $S^3$ for Chern-Simons theory,
one might expect this ${\cal N}=4$ super-Yang-Mills theory to be described
by an open string theory. 

As was described at Strings 2003, one can construct an open
string theory for self-dual ${\cal N}=4$ $d=4$ super-Yang-Mills
which closely resembles the $d=10$ pure spinor formalism. Unlike
the recent string theories constructed in twistor space
\ref\wit{E. Witten,
{\it Perturbative Gauge Theory as a String Theory in
Twistor Space}, hep-th/0312171\semi
N. Berkovits, {\it An Alternative String Theory in Twistor Space
for N=4 Super-Yang-Mills}, hep-th/0402045\semi
A. Neitzke and C. Vafa, {\it N=2 Strings and the Twistorial
Calabi-Yau}, hep-th/0402128.}, this
string theory is constructed in $(x,\t)$ space and only describes
the self-dual super-Yang-Mills interactions. However, this string
theory contains bosonic spinor worldsheet variables which might
be related to the recent twistor-string constructions.

In section 2, the $N=2$ string in $d=(2,2)$ target-space is 
reviewed using the $N=4$ topological description which involves
constant bosonic spinors. In section 3, it is shown that these
bosonic spinors can be treated as worldsheet variables if the
$d=4$ target space is supersymmetrized to an ${\cal N}=4$ $d=4$
target superspace. In section 4, this string theory is related
to a dimensional reduction of the $d=10$ pure spinor formalism
of the superstring. In sections 5 and 6, vertex operators and
scattering amplitudes are computed and shown to correspond to
self-dual ${\cal N}=4$ $d=4$ super-Yang-Mills. And section 7 contains
conclusions and speculations.

\newsec{Review of Open N=2 String}

Although no open string theory (at the time of the Strings 2003 talk)
was known to describe ${\cal N}=4$ $d=4$ super-Yang-Mills theory, 
there is an open string theory which describes ${\cal N}=0$ self-dual
Yang-Mills theory in $d=(2,2)$. This string theory is described by
the worldsheet action \ref\adem{M. Ademollo, L. Brink. A. D'Adda, R. D'Auria,
E. Napolitano, S. Sciuto, E. Del Giudice, P. DiVecchia, S. Ferrara,
F. Gliozzi, R. Musto, R. Pettorino and J.H. Schwarz, {\it Dual String
with U(1) Color Symmetry}, Nucl. Phys. B111 (1976) 77.} 
\ref\oog{H. Ooguri and C. Vafa, {\it Selfduality and N=2 String Magic},
Mod. Phys. Lett. A5 (1990) 1389\semi H. Ooguri and C. Vafa,
{\it Geometry of N=2 Strings}, Nucl. Phys. B361 (1991) 469\semi
H. Ooguri and C. Vafa, {\it N=2 Heterotic Strings},
Nucl. Phys. B367 (1991) 83.}\ref\lizzi{
A. D'Adda and F. Lizzi, {\it Space Dimensions from Supersymmetry
for the N=2 Spinning String: A Four-Dimensional Model},
Phys. Lett. B191 (1987) 85.}\ref\marcus{N. Marcus, {\it
The N=2 Open String}, Nucl. Phys. B387 (1992) 263,
hep-th/9207024\semi N. Marcus, {\it
A Tour Through N=2 Strings}, hep-th/9211059.}
\eqn\nzero{S=\int d^2 z (\half \p x^{a\ad}\pb x_{a\ad} +
\eta^\ad \pb\psi_\ad + \bar\eta^\ad \p\bar\psi_\ad )}
and the $\hat c=2$ N=2 superconformal generators
\eqn\gen{ T = \half \p x^{a\ad} \p x_{a\ad} + \half (\eta^\ad \p\psi_\ad
- \psi^\ad \p\eta_{\ad}) ,} 
$$G^+ = \psi^\ad \p x_{1 \ad}, \quad G^- = \eta^\ad \p x_{2 \ad},$$
$$ J = \eta^\ad \psi_\ad,$$
where $a$ and $\ad$ are independent two-component SL(2,R) indices and
$(\eta^\ad,\psi^\ad)$ and 
$(\bar\eta^\ad,\bar\psi^\ad)$ are left and right-moving fermions.
Physical states in N=2 string theory
are defined as dimension-zero U(1)-neutral
primary fields with
respect to these generators, and the only physical state is
$$\phi(x(z)) = e^{ik^{a\ad} x_{a\ad}(z)}$$
where $k^{a\ad} k_{a\ad}=0$. 

Computing the tree-level scattering
amplitudes of these physical states, one finds that $\phi(x)$ satisfies
the classical equations of motion \oog
\eqn\yang{\e^{\ad\bd} \p_{1\ad} (e^{-\phi} \p_{2\bd} e^\phi) =0,}
which describes a self-dual Yang-Mills field in the Yang description.
Note that in the Yang description, the self-dual gauge field $A_{a\ad}$
is defined non-covariantly in terms of $\phi$ as
\eqn\defaa{A_{1\ad}=0,\quad A_{2\ad} = e^{-\phi} \p_{2 \ad} e^\phi,}
which implies that the anti-self-dual field strength
\eqn\fab{F_{ab} = \e^{\ad\bd} (\p_{a\ad} A_{b\bd} -\p_{b\bd} A_{a\ad} + 
[A_{a\ad},A_{b\bd}])}
vanishes when $\phi$ satisfies \yang.

Although the equation $F_{ab}=0$
is manifestly Lorentz covariant, one cannot obtain \yang\ from
a covariant action constructed from $\phi$. A related problem is
that loop amplitudes for $\phi(x)$
computed using the N=2 string theory do not correspond to
loop amplitudes of self-dual Yang-Mills theory \marcus.

To make Lorentz invariance manifest, it is useful to construct
a set of ``small'' N=4 superconformal generators out of the
worldsheet variables as \ref\siegel{W. Siegel, {\it The N=4 String is
the Same as the N=2 String}, Phys. Rev. Lett. 69 (1992) 1493,
hep-th/9204005.}\ref\topo{N. Berkovits and C. Vafa, {\it N=4
Topological Strings}, Nucl. Phys. B433 (1995) 123, hep-th/9407190.}
\eqn\superconf{ T =
 \half \p x^{a\ad} \p x_{a\ad} + \half (\eta^\ad \p\psi_\ad
- \psi^\ad \p\eta_{\ad}) ,} 
$$G^+ = \psi^\ad \p x_{1\ad}, \quad \tilde G^+ = 
\psi^\ad \p x_{2\ad},\quad
G^- = \eta^\ad \p x_{2\ad},\quad \tilde G^- = -\eta^\ad \p x_{1\ad},$$
$$J^{++} = \psi^\ad \psi_\ad, \quad J = \psi^\ad \eta_\ad,\quad
J^{--} = \eta^\ad \eta_\ad,$$
where ``small'' means there is an $SL(2,R)$ set of dimension-one currents
as opposed to an $SO(2,2)$ set. 
 
As shown in \topo, one can describe this N=2 string as an ``N=4 topological
string'' by twisting
$\psi^\ad$ to have spin zero 
and $\eta^\ad$ to have spin one, so that 
$(G^+,\tilde G^+)$
have spin one and 
$(G^-,\tilde G^-)$ have spin two. Since the stress tensor has zero
central charge after twisting, there is no need to introduce
super-Virasoro ghosts. 
Physical states are described by
the cohomology of the BRST operator 
\eqn\brst{Q = \l^1 \int dz G^+ + \l^2 \int dz\tilde G^+ = \l^a
\int dz \psi^\ad
\p x_{a\ad}}
where $(\l^1,\l^2)$ are constants which parameterize
the coset
$SL(2,R)/GL(1)$.
This is a natural N=4 generalization of the 
N=2 topological string where physical states are described by the cohomology
of the
BRST operator $Q=\int dz G^+$. 

One finds that the unique state in the cohomology
of \brst\ is 
$$V = \l^a \psi^\ad A_{a\ad}(x)$$
where 
$ \e^{\ad\bd} (\p_{a\ad} A_{b\bd} -\p_{b\bd} A_{a\ad} )=0$. So the
vertex operator
$V$ describes self-dual Yang-Mills in a Lorentz-covariant manner.
However, using this N=4 topological description,
tree amplitudes involving $V$ are not
manifestly Lorentz covariant and loop amplitudes 
do not correspond to those of self-dual Yang-Mills theory. 

\newsec{String Theory for Self-Dual Super-Yang-Mills}

The lack of manifest Lorentz covariance is related to the fact that
$\l^a$ is a constant parameter and not a worldsheet variable. So
to restore Lorentz invariance \foot{Another approach to restoring
Lorentz invariance is to introduce Chan-Paton factors which describe
different spacetime helicities and imply spacetime supersymmetry
\ref\sieg{W. Siegel, {\it The N=2(4) String is Selfdual N=4
Yang-Mills}, hep-th/9205075.}
\ref\chalmers{G. Chalmers and
W. Siegel, {\it Global Conformal Anomaly in N=2 String},
Phys. Rev. D64 (2001) 026001, hep-th/0010238.}. See also \ref\olaf{
A. Galajinsky and O. Lechtenfeld, {\it Towards a Stringy Extension
of Selfdual Super-Yang-Mills}, Phys. Lett. B460 (1999) 288, hep-th/9903149\semi
S. Belluci and A. Galajinsky, {\it
Restoring Lorentz Invariance in Classical N=2 String},
Nucl. Phys. B606 (2001) 119, hep-th/0104003\semi
S. Belluci, A. Galajinsky and O. Lechtenfeld, {\it
A Heterotic N=2 String with Spacetime Supersymmetry}, Nucl. Phys.
B609 (2001) 410, hep-th/0103049\semi
J. Bischoff, S. Ketov and O. Lechtenfeld, {\it
GSO Projections, BRST Cohomology and Picture-Changing in N=2
String Theory}, Nucl. Phys. B438 (1995) 373, hep-th/9406101\semi
Z. Khviengia, H. Lu, C.N. Pope, E. Sezgin, X.J. Wang and K.W. Xu,
{\it N=1 Superstring in (2+2) Dimensions},
Nucl. Phys. B444 (1995) 468, hep-th/9504121\semi
H. Lu, C.N. Pope and E. Sezgin,
{\it A Search for New (2,2) Strings},
Class. Quant. Grav. 12 (1995) 1913,
hep-th/9504122\semi
J. deBoer and K. Skenderis, {\it
Self-dual Supergravity from N=2 Strings},
Nucl. Phys. B500 (1997) 192,
hep-th/9704040.} for other approaches to restoring
Lorentz invariance and supersymmetry.}, one should treat $\l^a$ as a worldsheet
variable and define $Q=\int dz \l^a\psi^\ad \p x_{a\ad}$ . But since 
$$G^+(y) \tilde G^+ (z) \to {{J^{++}(y) + J^{++}(z)}\over {2(y-z)^2}},$$
$Q^2 $
is nonvanishing when $\l^a$ is not constant and one needs to modify
the BRST charge. One possible modification is 
$$Q = \int dz (\l^1 G^+ + \l^2 \tilde G^+ + e J^{++} + h J + {\rm
~~~ghost~~terms})$$
\eqn\modif{=\int dz (\l^a \psi^\ad \p x_{a\ad}
+ e \psi^\ad \psi_\ad + f \l^a \p \l_a + h(-\l^a w_a +\psi^\ad \eta_\ad
+2 fe))}
where $(w_a, f,g)$ are the conjugate antighosts for $(\l^a, e,h)$.
The Lorentz covariant worldsheet action for these fields is
\eqn\waction{S =\int d^2 z(\half \p x^{a\ad} \pb x_{a\ad}
+ \eta^\ad \pb \psi_\ad + w^a \pb \l_a + e \pb f + g \pb h
+ \bar\eta^\ad \p \bar\psi_\ad +\bar w^a \p\bar \l_a +\bar e \p\bar f +
\bar g \p\bar h),}
and the desired effect of the fermionic ghosts $(e,f)$ and
$(g,h)$ is
to cancel the contribution
of the bosonic $(w_\a,\l^\a)$ ghosts to the BRST cohomology.
This is plausible since
the $h$ ghost couples in $Q$ to the constraint
$J= -\l^a w_a +\psi^\ad \eta_\ad + 2fe$ which generates
projective transformations of $\l^a$, and the $f$ ghost couples to
the constraint $\l^a \p \l_a$ which fixes $\l^2/\l^1$ to be constant.
However, since $J(y) J(z)\to 4 (y-z)^{-2}$, 
$Q^2 = 4\int dz (h \p h)$ and the theory is anomalous.
Fortunately, as will now be shown, this anomaly can be cured by
replacing ${\cal N}=0$ self-dual Yang-Mills with ${\cal N}=4$ self-dual
super-Yang-Mills. 

As shown by Siegel in \sieg\ref\gsdual{W. Siegel,
{\it Green-Schwarz Formulation of Self-Dual Superstring},
Phys. Rev. D47 (1993) 2512, hep-th/9210008.}, 
one can describe self-dual super-Yang-Mills
in superspace by extending
the $x^{a\ad}$ spacetime variables to $(x^{a\ad},\t^{a j})$
superspace variables where $j=1$ to $\cal N$. Note that one does
not need to include $\bar\theta^{\ad\bar j}$ variables since the antichiral
superspace derivatives $\bar\nabla_{\ad\bar j}$ satisfy
the trivial (anti)commutation relations 
$$\{\bar\nabla_{\ad\bar j},
\bar\nabla_{\bd\bar k} \}=[
\bar\nabla_{\bd\bar k}, \nabla_{a\ad}]=0$$
when
the theory is self-dual. It will be convenient to combine the 
$(x^{a\ad},\t^{a j})$ variables into a superspace variable
\eqn\Ydef{Y^{a J} = (x^{a\ad}, \t^{a j})}
where $J = (j,\ad)$ is an $OSp({\cal N}|2)$ vector index.
The variable $Y^{a J}$ is defined to transform covariantly under
$SL(2,R)$ rotations of the $a$ index and under $OSp({\cal N}|2)$ rotations
of the $J$ index. Just as $(\eta^\ad,\psi^\ad)$ are the fermionic
worldsheet superpartners of $x^{a\ad}$, one can define $(b^j, c^j)$
to be the bosonic worldsheet superpartners of $\t^{a j}$. These
worldsheet 
superpartners of $Y^{a J}$
can be described in $OSp({\cal N}, 2)$ notation as
\eqn\Cdef{C^J = (\psi^\ad,c^j),\quad B^J = (\eta^\ad, b^j).}

The open string for self-dual super-Yang-Mills will be defined by
the worldsheet action
\eqn\saction{S =\int d^2 z(\half \p Y^{a J} \pb Y_{a J}
+ B^J \pb C_J + w^a \pb \l_a + e \pb f + g \pb h
+ \bar B^J \p  \bar C_J +\bar w^a \p\bar \l_a +\bar e \p\bar f +
\bar g \p\bar h),}
the stress tensor 
\eqn\sstress{T=\half \p Y^{a J}\p Y_{a J} + B^J \p C_J
+ w^a \p \l_a + e\p f + g \p h,}
and the BRST operator
\eqn\smodif{Q=\int dz (\l^a C^J \p Y_{a J}
+ e C^J C_J + f \l^a \p \l_a + h(-\l^a w_a + C^J B_J
+2 fe)),}
where the $J$ index is raised and lowered using the 
$OSp({\cal N}| 2)$ metric which is symmetric/antisymmetric
when $J=j/\ad$. Note that the action is invariant 
under only an $SL(2)\times OSp({\cal N}|2)$ subgroup of 
the full superconformal group $PSU(2,2|4)$. Also note that
the kinetic term is
$\half\int d^2 z \p Y^{a J}\pb Y_{a J}= \half\int d^2 z (
\p x^{a\ad} \pb x_{a\ad} + \p\t^{a j} \pb \t_{a j}),$
so $\t^{a j}$ satisfies the non-holomorphic OPE's 
$$\t^{a j}(y,\bar y) \t^{b k}(z,\bar z) \to \e^{ab} \d^{jk} \log |y-z|^2.$$
One can easily check that $J= -\l^a w_a + C^J B_J + 2 fe$ now satisfies
the OPE $J(y) J(z) \to (4-{\cal N}) (y-z)^{-2}$, so the theory is
not anomalous when ${\cal N}=4$. 

\newsec{Comparison with Pure Spinor Formalism}

To show that the theory described by \saction, \sstress\ and \smodif\
describes ${\cal N}=4$ self-dual super-Yang-Mills, it is useful to
first review the pure spinor formalism of the superstring in a flat
$d=10$ background \pure. In a flat $d=10$ background, the pure spinor
formalism for the superstring is described by the worldsheet action
\eqn\paction{S =\int d^2 z (\half \p x^m \pb x_m +
p_\a \pb \t^\a + w_\a\pb \l^\a +
\bar p_\a \p \bar\t^\a + \bar w_\a\p\bar \l^\a ),}
the stress tensor
\eqn\pstress{T =\half \p x^m \p x_m + p_\a \p\t^\a + w_\a \p\l^\a,}
and the BRST operator 
\eqn\pbrst{Q= \int dz \l^\a d_\a}
where $m=0$ to 9, $\a=1$ to 16, 
$$d_\a = p_\a +  \g^m_{\a\b}\p x_m \t^\b -\half (\t\g^m\p\t)(\g_m\t)_\a$$ 
is the Green-Schwarz
constraint, and $\l^\a$ is a pure spinor satisfying
$$\l^\a \g^m_{\a\b} \l^\b =0.$$

Massless $d=10$ super-Yang-Mills states of the open superstring are described
by the unintegrated vertex operator
\eqn\punint{V= \l^\a A_\a (x,\t)}
and by the integrated vertex operator
\eqn\pint{U=\p \t^\a A_\a (x,\t) + \pi^m A_m (x,\t) +
d_\a W^\a(x,\t) + \half(w\g^{mn}\l) F_{mn}(x,\t)}
where $\pi^m=\p x^m +\t\g^m\p\t$ is the supersymmetric
momentum, $(A_\a,A_m)$ are the spinor and vector gauge superfields satisfying
\eqn\vec{D_{(\a} A_{\b)}= \g^m_{\a\b} A_m,}
and $W^\a$ and $F_{mn}$ are the
spinor and vector field strength superfields. Although there is no $b$
ghost in this formalism, the integrated and unintegrated vertex operators
are related to each other by $QU = \p V$. 

$N$-point tree amplitudes
are computed in the pure spinor formalism by the correlation function
\eqn\treea{A= \langle V_1(z_1) V_2(z_2) V_3 (z_3) \int dz_4 U_4 (z_4) ...
\int dz_N U_N(z_N)\rangle }
on a disk.
Since the worldsheet action is quadratic, the only nontrivial aspect of
computing this correlation function is the normalization of worldsheet
zero modes. The correct normalization is 
\eqn\zerom{\langle (\l \g^m\t)(\l\g^n\t)(\l\g^p\t)(\t\g_{mnp}\t)
\rangle =1,} 
which is BRST invariant since
$(\l \g^m\t)(\l\g^n\t)(\l\g^p\t)(\t\g_{mnp}\t)$
is the unique state of ghost-number three in the BRST cohomology.
Recall that in bosonic string theory, $c\p c\p^2 c$ is the unique state
of ghost-number three in the BRST cohomology and $\langle c \p c 
\p^2 c\rangle =1$ is the correct normalization for zero modes on a disk.
Furthermore, one can check that the field theory action
\eqn\sft{{\cal S}=\langle VQV + V^3\rangle}
reproduces the standard $d=10$ 
super-Yang-Mills action if one uses the normalization
of \zerom\ and the string field $V=\l^\a A_\a (x,\t)$.

Since $d=10$ super-Yang-Mills dimensionally reduces to 
${\cal N}=4$ $d=4$ super-Yang-Mills (which has the same spectrum as
${\cal N}=4$ self-dual super-Yang-Mills), it is natural to expect
that the open string theory for 
${\cal N}=4$ self-dual super-Yang-Mills is related to a dimensional
reduction of the $d=10$ pure spinor formalism. This relationship can
be made more explicit by first decomposing the $d=10$ spinors
$\l^\a $ and $\t^\a$ for $\a=1$ to 16 into the $d=4$ $SU(4)$ spinors
$(\l^{a j},\bar\l^{\ad \bar j})$
and $(\t^{a j},\bar\t^{\ad \bar j})$ where $(a,\ad)=1$ to 2 and
$(j,\bar j)=1$ to 4. If one then sets\foot{A similar decomposition
of pure spinors has recently been used in the interesting paper
of \ref\grassi{P.A. Grassi and P. van Nieuwenhuizen, {\it
Harmonic Superspaces from Superstrings},
hep-th/0402189.} to obtain $d=4$ harmonic superspaces with ${\cal N}=2,3$ or 4
supersymmetry.}
\eqn\sets{\l^{a j} = \l^a c^j, \quad \bar\l^{\ad\bar j} =0,}
the pure spinor constraint is satisfied and the BRST operator of 
\pbrst\ becomes
\eqn\newq{Q = \int dz \l^{a j} d_{a j} = \int dz \l^a c^j
(p_{a j} + \bar\t^{\ad\bar j} \p x_{a\ad} + ...),}
where $...$ includes terms
quadratic in $\bar\t^{\ad \bar j}$.
After defining 
\eqn\psidefin{\psi^\ad = c^j \bar\t^{\ad\bar j}}
and recognizing that $\p\t^{a j}$ in the second-order action of \saction\ 
plays the role of $p_{a j}$ in the first-order action of \paction, one can
reexpress the BRST operator of \newq\ as
\eqn\newbrst{Q = \int dz \l^a (c^j \p\t_{a j} + \psi^\ad \p x_{a\ad} + ...),}
which closely resembles \smodif. 
In the same sense that the pure spinor formalism can be understood
as covariant quantization of the Green-Schwarz superstring, the
self-dual formalism described by \smodif\
can be understood as covariant quantization of
the self-dual Green-Schwarz superstring proposed by Siegel in \gsdual. 
 
\newsec{Self-Dual Super-Yang-Mills Vertex Operators}

By analyzing the cohomology of the BRST operator
of \smodif, one finds a similar structure to
the massless cohomology in the pure spinor formalism. At ghost number one
(where the ghost-number charge is defined as 
$\int dz(\l^a w_a + e f - gh)$), the unique state
in the cohomology is described by the unintegrated vertex operator 
\eqn\unint{V = \l^a C^J A_{a J}(Y) = \l^a (\psi^\ad A_{a\ad}(x,\t) +
c^j A_{a j}(x,\t)),}
where $A_{a J}$ satisfies
\eqn\eoms{\p_{a J} A_{b K} - (-1)^{s(K)} \p_{b K} A_{a J} =\e_{ab} F_{JK}}
and $s(K)= 1/0$ when $K$ is $j/\ad$. Although there is no $b$ ghost, one
can define the integrated vertex operator by $QU= \p V$ and one finds that
\eqn\integ{U = \p Y^{a J} A_{a J}(Y) + B^J C^K F_{JK}(Y).}
Note the close resemblance of equations \unint - \integ\ to equations
\punint - \vec\ in the pure spinor formalism. 

To see that these vertex operators
describe self-dual super-Yang-Mills, note that \eoms\
is the supersymmetrization of the ${\cal N}=0$ self-dual Yang-Mills
relation 
$[\nabla_{a \ad}, \nabla_{b \bd}] = \e_{ab} F_{\ad \bd}$. 
In components,
$A_{a J}$ can be gauge-fixed to
\eqn\comp{A_{a\ad} = a_{a\ad} + \t^j_a \bar \xi_{\ad j} +
\t^j_a \t^{b k} \p_{b \ad} \phi^{lm} \e_{jklm}
+ \t_a^j\t^{b k} \t^{c l}\p_{b\ad} \xi^m_c \e_{jklm}
+(\t^4)^{bcd}_a \p_{b \ad} G_{cd},}
$$A_{a j} = (\t^k_a \phi^{lm} +\t_a^k \t^l_b \xi^{b m} +
\t^k_a \t^l_b\t^m_c G^{bc})\e_{jklm},$$
where $a_{a\ad}$ is the self-dual gauge field, $\xi^{b m}$
and $\bar \xi_{\ad j}$ are the on-shell
gluinos, $\phi^{lm}$ are the six on-shell scalars,
and $G^{bc}$ is the anti-self-dual field strength. The ${\cal N}=4$
self-dual super-Yang-Mills action 
for these component fields is
\eqn\actsd{S = Tr \int d^4 x[( \p_{a\ad} a_b^\ad + \p_{b\ad} a_a^\ad
+ [a_{a\ad},a_b^\ad]) G^{ab} +
\xi^{b m} \nabla_{b\bd} \bar\xi_m^\bd + 
\nabla_{a\ad} \phi^{jk} \nabla^{a\ad} \phi^{lm} \e_{jklm}],}
where $\nabla_{a\ad} = \p_{a\ad} + a_{a\ad}$. Note that this action
has the same spectrum as ordinary ${\cal N}=4$ super-Yang-Mills but
has different interactions \sieg.

Unlike the pure spinor formalism, there are no massive states in the
cohomology of $Q$ of \smodif. As in the ${\cal N}=0$ self-dual string, this
is because the constraints $C^J \p Y_{a J}$ in the BRST operator
imply that all states in the cohomology are independent of the nonzero
worldsheet modes of $Y_{a J}$ and $(C^J, B_J)$. Furthermore, states
are independent of nonzero modes of $(\l^a,w_a,e,f,g,h)$ because of
the terms $f \l^a \p \l_a$ and $h(-\l^a w_a + C^J B_J +2ef)$ in $Q$.
The absence of physical massive states will be verified below
by showing there are no massive poles in the scattering amplitudes.

\newsec{Self-Dual Super-Yang-Mills Tree Amplitudes}

To compute $N$-point tree amplitudes in this formalism, one uses the
same prescription as before that 
\eqn\trees{A= \langle V_1(z_1) V_2(z_2) V_3 (z_3) \int dz_4 U_4 (z_4) ...
\int dz_N U_N(z_N)\rangle. }
Since the unique state at ghost-number three in the cohomology is
\eqn\uniq{\l^a \l^b \l^c \psi^\ad \psi_\ad \t^j_a \t^k_b \t^l_c c^m
\e_{jklm},} 
one should define normalization of the zero modes by
\eqn\zeroq{\langle
\l^a \l^b \l^c \psi^\ad \psi_\ad \t^j_a \t^k_b \t^l_c c^m
\e_{jklm}\rangle =1.} 
Although this normalization appears strange, it is obtained from the
$d=10$ normalization 
$\langle(\l\g^m\t)(\l\g^n\t)(\l\g^p\t)(\t\g_{mnp}\t)
\rangle =1$ by performing the dimensional reduction of \sets\
and
\psidefin\ where
one sets
$\l^{a j}=\l^a c^j,$
$\bar\l^{\ad\bar j} =0,$ and $\psi^\ad = c^j \bar\t^{\ad\bar j}$.

One can easily check that this tree amplitude prescription reproduces
the desired self-dual super-Yang-Mills amplitudes, which vanish unless
the external momenta are chosen to lie in the same ``self-dual plane''.
To see this, note that when the self-dual gauge superfield $A_{a J}(Y)$
is onshell, it can be expressed as
\eqn\expr{A_{a\ad}(Y)= \int d^4 \k ~~\bar\pi_\ad e^{i \pi^b
(\bar\pi^\bd x_{b\bd} + \k^j \t_{b j}) }\Phi_a(\pi,\bar\pi,\k),}
$$A_{a j}(Y)= \int d^4 \k~~  \k_j e^{i \pi^b
(\bar\pi^\bd x_{b\bd} + \k^j \t_{b j})} \Phi_a(\pi,\bar\pi,\k),$$
where $ \Phi_a(\pi,\bar\pi,\k)$ is unconstrained,
$p_{a\ad} =\pi_a \bar\pi_\ad$ is the momentum, and 
$\k_j$ is the fermionic conjugate momentum to $\pi^a \t_{a j}$.
These superfields can be expressed in $OSp(4|2)$-covariant notation as
\eqn\expoo{A_{a J}(Y)= \int d^4 \k ~~ \bar\Pi_J e^{i \pi^b
\bar\Pi^J Y_{b J} } \Phi_a(\pi,\bar\Pi),}
where $\bar\Pi^\ad = \bar\pi^\ad$ and $\bar\Pi^j = \k^j$.
Since 
$$\p_{a J}A_{b K}= i\int d^4 \k ~~\pi_a\bar\Pi_J \bar\Pi_K e^{i \pi^c
\bar\Pi^L Y_{c L} } \Phi_b(\pi,\bar\Pi),$$
$A_{a J}$ of \expoo\ satisfies \eoms\ where
$$F_{JK}=
i\int d^4 \k ~~\bar\Pi_J \bar\Pi_K e^{i \pi^c
\bar\Pi^L Y_{c L} } \pi^a\Phi_a(\pi,\bar\Pi).$$
So the onshell unintegrated vertex operator is
\eqn\onshellV{V = \l^a C^J A_{a J}(Y) = \int d^4 \k (C^J \bar\Pi_J)
(\l^a \Phi_a(\pi,\bar \Pi))
\exp(i\pi^b \bar\Pi^J Y_{b J}).}

To show that scattering amplitudes of these vertex operators
satisfy the integrability properties of self-dual
Yang-Mills, note that
$$Q( e^{i\pi^b \bar\Pi^J Y_{b J}}) =i (\l^a \pi_a)(C^J \bar\Pi^J)
( e^{i\pi^b \bar\Pi^J Y_{b J}})$$
where $Q = \int dz (\l^a C^J \p Y_{a J} + ...)$.
So $V$ can formally
be written as 
\eqn\formal{V = Q [ (\l^a \pi_a)^{-1} \Omega ]}
where 
$$\Omega = -i \int d^4 \k
(\l^a \Phi_a(\pi,\bar \Pi))
\exp(i\pi^b \bar\Pi^J Y_{b J}).$$
Since $(\l^a \pi_a)^{-1}$ has singularities when $\l^a$ is
proportional to $\pi^a$, $(\l^a \pi_a)^{-1}$ is not globally defined
and $V$ is not BRST-trivial. However, the product of two onshell
vertex operators can be expressed either as 
\eqn\twoex{V_1 V_2  = Q [(\l^a \pi_{1 a})^{-1} \Omega_1 V_2] \quad
{\rm or}\quad 
V_1 V_2  = Q [(\l^a \pi_{2 a})^{-1} V_1 \Omega_2] .}
As long as $\pi_{1 a} \pi_2^a$ is nonzero, either 
$(\l^a \pi_{1 a})^{-1}$ or
$(\l^a \pi_{2 a})^{-1}$ is nonsingular. So whenever
$\pi_{1 a} \pi_2^a$ is nonzero, $V_1 V_2$ is BRST-trivial. This
implies that $N$-point tree amplitudes vanish unless 
$\pi_r^a \pi_{s a}=0$ for all $1\leq r,s\leq N$, i.e. unless all
external momenta are in the same ``self-dual plane''.
This is consistent with the absence of massive states in the spectrum
since $\pi_r^a \pi_{s a}=0$ implies that
all intermediate states are massless.

Unlike amplitudes computed using the ${\cal N}=0$ version of the
self-dual string, the amplitudes computed using this ${\cal N}=4$ 
formalism are manifestly Lorentz covariant. This does not cause
contradictions for loop amplitudes since all ${\cal N}=4$
self-dual loop amplitudes
are vanishing \sieg.
Furthermore, one can check that if one computes the field theory
action
\eqn\sft{{\cal S}= \langle VQV + V^3\rangle}
using the normalization of \zeroq\ and the string field
$V=\l^a C^J A_{a J} (Y)$, one reproduces the self-dual super-Yang-Mills
action of \actsd.

\newsec{Conclusions and Speculations}

In these proceedings, an open string theory for ${\cal N}=4$
$d=4$ self-dual super-Yang-Mills was constructed which
is related to the $d=10$ pure spinor formalism of the superstring.
Since
the coupling constant of self-dual (super)-Yang-Mills can be eliminated
by scaling $G^{ab}$ and $a_{a\ad}$ of \actsd\  in opposite directions, 
self-dual (super) Yang-Mills is essentially a free theory. However,
one can obtain 
the full (super) Yang-Mills action 
from the self-dual (super) Yang-Mills action of \actsd\ by adding
the perturbation term 
\eqn\pert{S_{pert} = g_{YM}\int d^4 x G^{ab} G_{ab} ~~~(+~~~{\rm 
supersymmetric~~terms})}
where $g_{YM}$ is the Yang-Mills coupling constant and
$G^{ab}$ is the anti-self-dual field strength of
\comp. 
It is therefore important to understand the perturbation of 
\pert\ which turns self-dual (super) Yang-Mills into ordinary
(super) Yang-Mills.\foot{ At the Strings 2003 talk, it was speculated
that this perturbation might be related to deforming the $R^4$ background
to an $AdS_4$ background. In an $AdS_4$ background, the $SL(2,R)$ and
$OSp(4|2)$ rotations of $Y_{a J}$ combine with the $R^{4|8}$ 
translations of $Y_{a J}$ to form a supergroup $OSp(4|4)$.
Since the $OSp(4|4)$ supergroup treats chiral and antichiral spinor
indices in a symmetric manner, it was hoped that the string theory
in this $AdS_4$ background might describe the full ${\cal N}=4$
super-Yang-Mills theory. However, there is a simple argument which
makes this speculation unlikely. If one performs an $O(4)$ $R$-symmetry
transformation which is not in the $SO(4)$ subgroup of $OSp(4|2)$,
the self-dual super-Yang-Mills action of \actsd\ 
transforms with opposite sign from the $g_{YM}\int d^4 x G^{ab} G_{ab}$
perturbation term. But since the term in the worldsheet action
which deforms $R^4$ to
$AdS_4$ is invariant under these $O(4)$ transformations, it
cannot be responsible for this change in sign.}

Recently, it has been understood how to describe this perturbation
using string theories in twistor space \wit.
It would therefore be very
interesting to relate these string theories in twistor space with
the string theory in $(x,\theta)$ space which has been described here.
Such a relationship would not be surprising
since the pure spinor formalism for the superstring is closely
related to theories in ten-dimensional twistor space \ref\tentw
{E. Witten, {\it Twistor-like Transform in Ten Dimensions},
Nucl. Phys. B266 (1986) 245\semi
D. Sorokin, V. Tkach, D. Volkov and A. Zheltukhin, {\it From 
the Superparticle Siegel Symmetry to the Spinning Particle
Proper Time Supersymmetry}, Phys. Lett. B216 (1989) 302\semi
N. Berkovits, {\it Calculation of 
Green-Schwarz Superstring Amplitudes using the N=2 Twistor-String
Formalism}, Nucl. Phys. B395 (1993) 77, hep-th/9404162\semi
P. Howe, {\it Pure Spinor Lines in Superspace and Ten-Dimensional
Supersymmetric Theories}, Phys. Lett. B258 (1991) 141\semi
M. Matone, L. Mazzucato, I. Oda, D. Sorokin and M. Tonin,
{\it The Superembedding Origin of the Berkovits Pure Spinor Covariant
Quantization of Superstrings}, Nucl. Phys. B639 (2002) 182,
hep-th/0206104.}.
Furthermore, equations \expr - \twoex\ for vertex operators and
scattering amplitudes in this string theory resemble twistor
constructions involving the Penrose twistor-transform.

\vskip 15pt

{\bf Acknowledgements:}
I would especially
like to thank Edward Witten for many useful ideas and comments and
for his collaboration during various stages of this project. I would
also like to thank Rajesh Gopakumar, Warren Siegel and Cumrun Vafa for
useful discussions,  
CNPq grant 300256/94-9, Pronex grant 66.2002/1998-9, 
and FAPESP grant 99/12763-0 for partial financial support, 
the Institute for Advanced Study for their hospitality, and
the organizers of Strings 2003 for their excellent work.

\listrefs

\end